\documentclass[12pt,a4paper]{article}

\usepackage[T1]{fontenc} 
\usepackage{feynmf}
\usepackage{graphics}
\usepackage{graphicx}
\newcommand{\begeq}{\begin{equation}} 
\newcommand{\fineeq}{\end{equation}} 
\newcommand{\begar}{\begin{eqnarray}} 
\newcommand{\finear}{\end{eqnarray}}

\begin{document} 
     
\begin{center}
\Large \bfseries
The see-saw mechanism  \\and \\heavy Majorana neutrino masses \\in an SO(10)
model \\[0.5cm]
\large \mdseries \upshape
W. Alles\footnote{e-mail: alles@bo.infn.it} and L. Frassinetti
\footnote{e-mail: frasso\_l@yahoo.it}

$^{1,2}$ Dipartimento di Fisica, Universit\'{a} di Bologna, Italy

$^1$ INFN, Sezione di Bologna
\\[0.5cm]
Abstract
\\[0.25cm]
\end{center}
\normalsize
We apply the see-saw mechanism and a SO(10) model to neutrino masses and
mixing in order to estimate the heavy Majorana
masses. We discuss shortly the decay modes of heavy Majorana neutrinos
and calculate their contribution to the lepton number violating processes
$\mu\rightarrow e\gamma$, $\tau\rightarrow \mu\gamma$ and $\tau\rightarrow
e\gamma$. 
\section{Introduction}
The Standard Model fermionic fields do not include the right-handed neutrino
field and so it is impossible to have a Dirac mass term for this particle
provided it is not introduced {\it ad hoc}. In this case we will have the
following lagrangian mass term:\hspace{0.5cm}  $\mathcal{L}=f_\nu
\overline{l}_L \nu_R  \phi+h.c.\rightarrow m_D \overline{\nu_L}\nu_R+h.c.$ , 
where $m_D= \frac{1} {\sqrt{2}}vf_\nu $ is the Dirac mass. To have a neutrino
mass of order of {\it eV} we should have a Yukawa coupling $f_\nu\sim 10^{-11}$
which seems very unlikely.

The best explanation of the lightness of neutrino comes from the
see-saw mechanism \cite{seesaw}.This mechanism works in a natural way  
in the framework of grand unification theories such as $SO(10)$  but 
also in the Standard Model, if we include a right-handed neutrino
field. The mechanism is based on three simple hypotheses, 
motivated in the grandunification model: 1) a zero Majorana mass 
for left-handed neutrinos ($m_L\simeq 0$), 2) the neutrino Dirac mass 
should be comparable to the charged fermion masses ($m_D\simeq m^f$) and 3) the
 Majorana mass of $\nu_R$  should be much larger than the Dirac ones
($m_R\simeq\mathcal{M}\gg m_D$). In this way the lagrangian mass terms, in the
one generation case, will be: \begeq 
   \mathcal{L}^{D+M}=-m_D\overline 
   {\nu_R}\nu_L-\frac{1}{2}m_R\overline{\nu_R} \nu_R^c+h.c.=-m^f\overline 
   {\nu_R}\nu_L-\frac{1}{2}\mathcal{M}\overline{\nu_R} \nu_R^c+h.c. 
   =-\frac{1}{2}\overline{n_L^c}M n_L+h.c. 
\fineeq 
with:
\begeq 
   M\equiv\left(\begin{array}{cc} 0 & m^f \\ m^f & \mathcal{M}\end{array}\right)  
\fineeq  
The eigenvalues of this mass matrix will be the neutrino masses: 
\begeq
  m_1  \simeq  -\frac{(m^f)^2}{\mathcal{M}} \ll m^f  \qquad
  m_2  \simeq  \mathcal{M} 
\fineeq
obtaining a very low mass, which would explain the lightness of neutrino, and a
very high mass, for a superheavy neutrino. 

In the three generations case we will have a $6\times 6$ mass matrix:
\begeq 
 M=\left(\begin{array}{cc} 0 & M_D^T \\ M_D & M_R \end{array}   
   \right)  
\fineeq  
where $M_D$ and $M_R$ are $3\times3$ matrixes ($M_R$ is also symmetric). We may
block diagonalize $M$ by a unitary trasformation \cite{bilbeta}, obtaining the
$3\times 3$ mass matrixes for light and heavy neutrinos \cite{Kanaya,ak}:
\begar 
   M_{light}&\simeq& -M_D^T M_R^{-1}M_D\\ 
   M_{heavy} &\simeq& M_R 
\finear
Thus heavy neutrino masses are the eigenvalues of the right-handed
Majorana mass matrix.

\section{Mixing of quarks and leptons in SO(10)}

The mixing of fermions derives from the fact that the mass matrix is not
diagonal and so the weak eigenstates are different from the mass eigenstates.
In the quark sector the mass matrixes of up quarks $M_u$ and of down quarks
$M_d$ may be diagonalized by a biunitary transformation: 
\begar 
  L_u^\dagger M_u R_u &=& M^{diag}_u  \nonumber\\  
  L_d^\dagger M_d R_d &=& M^{diag}_d  
\finear 
and thus the Cabibbo-Kobayashi-Maskawa matrix is:
\begeq 
  K=L_u^\dagger L_d 
\fineeq 
and the mixing of quarks is:
\begeq 
  \left(\begin{array}{c} d'\\s'\\b' \end{array}\right)=K   
  \left(\begin{array}{c} d\\s\\b \end{array}\right) 
\fineeq 
where the weak eigenstates are labelled by $'$.
In the same way, in the leptonic sector we have:
\begar 
  L_\nu^\dagger M_{\nu} R_\nu &=& M^{diag}_{\nu}  \nonumber\\  
  L_l^\dagger M_l R_l &=& M^{diag}_l  
\finear 
where $M_\nu$ is the mass matrix of light neutrinos and $M_l$ is the mass
matrix of charged leptons; the mixing matrix of neutrinos is:
\begeq 
  U^\dagger=L_\nu^\dagger L_l 
\fineeq 
and thus we obtain: 
\begeq    
  \left(\begin{array}{c} \nu_e\\ \nu_\mu\\ \nu_\tau \end{array}\right)=U   
  \left(\begin{array}{c} \nu_1\\ \nu_2\\ \nu_3 \end{array}\right) 
\fineeq 
The problem is that it is the $6\times 6$ mass matrix (4) that should be
diagonalized not  the mass matrix of light neutrinos. Yet it can be
shown \cite{tesi}  that the same result is achieved if we
consider the $3\times 3$ mass matrix of light neutrinos.

The 16 left-handed fields of one generation of fermions transforms like a
16-component SO(10) spinor. The product $16\otimes 16= 10\oplus 120\oplus 126$
leaves open the possibility for a 10-, a 120- or a 126-Higgs multiplet. The
only SU(5) singlet is contained in 126 and leads to a very large Majorana mass
for $\nu_R$. Therefore we exclude the possibility of  SU(2) doublet Higgs
contained in 126 and limit our choice, for simplicity, to the Higgs mesons of
representation 10.

In the SO(10) model \cite{georgi,fritzsch,lan}, using a Higgs
decuplet, we obtain symmetric mass matrix and $M_d=M_l $ and $M_u=M_D$, where
$M_d$ and $M_u$ are the mass matrix of {\it down} and {\it up} quarks, $M_l$
the mass matrix of charged leptons and $M_D$ the Dirac mass matrix of neutrino
\cite{moa,moagen,moamassa}. Thus the mass matrix of light neutrino will be
$M_\nu = M_u M^{-1}_R M_u$. As a consequence we obtain $L_l=L_d$ and for the
matrix $L_\nu$ we can write: 
\begar   
M_\nu &=& M_u M^{-1}_R M_u \nonumber\\ 
 &=& L_u M_u^{diag} R_u^\dagger M^{-1}_R L_u M_u^{diag} R_u^\dagger 
\finear 
Defining $M_1=M_u^{diag} R_u^\dagger M^{-1}_R L_u M_u^{diag}$ and 
calling $A$ and $A_1$ the matrixes which diagonalize it, we obtain: 
\begar 
  M_\nu&=&L_u M_1   R_u^\dagger \nonumber\\ 
       &=& L_u A M_\nu^{diag} A_1^\dagger   R_u^\dagger \\ 
  M_\nu &=& L_\nu M_\nu^{diag} R_\nu^\dagger 
\finear 
Thus we may assume: 
\begeq 
  L_\nu=L_u A \qquad L_\nu^\dagger=A^\dagger L_u^\dagger 
\fineeq 
and so: 
\begeq 
 \fbox{$\displaystyle U^\dagger=A^\dagger L_u^\dagger L_l=A^\dagger K$} 
\fineeq 
We think  the importance of  Majorana mass
matrix of right-handed neutrinos is worth being stressed and consequently of
the heavy Majorana neutrinos. As a matter of fact, in their absence the see-saw
mechanism cannot be used and besides having equality between neutrino and
quark masses, we obtain equality between their mixing, which is experimentally
unacceptable.

Moreover,  equation  (17) can show that the quadratic and
linear see-saw models \cite{byl} are not correct \cite{tesi}. In the  quadratic
see-saw the Majorana mass matrix of right-handed neutrinos is taken as
proportional to the identity matrix and all the heavy neutrinos have  the same
mass. In this case we can see that the matrix $A$ is equal to the identity
matrix and therefore, even if we can use the see-saw mechanism, we obtain that
the mixing of quarks and neutrinos is the same. Thus we can say that not all
the heavy neutrinos have the same mass and that the quadratic see-saw is not
correct. Likewise it is possible to see that also the linear see-saw is not
correct, so the Majorana mass matrix of right-handed neutrinos is different
from the mass matrix of up quarks.    

\section{Masses of heavy Majorana neutrinos}
In this section we are going to try to find a form for the mass matrix of the
heavy Majorana neutrinos ($M_R$) and consequently an estimate of their masses.

From eq. (6),(13) and (14) we obtain (assuming $M_u$ and $M_R$ real matrixes
and remembering they are simmetric):
\begeq 
 M_{heavy}\simeq M_R=L_u M_u^{diag} A  M_\nu^{diag-1} A^\dagger
M_u^{diag}R_u^\dagger  \fineeq 
The problem is that we do not know the form of $L_u$, $R_u$ and $A$. But if we
assume $K\simeq I$ and $L_u\simeq R_u\simeq I$ \cite{buccella} we have
$A\simeq U$: 
\begeq 
   M_{heavy}=M_u^{diag}U M_\nu^{diag-1} U^\dagger M_u^{diag}
\fineeq
The form of neutrino  mixing matrix depends on the kind of solution of
solar neutrinos. In this work we consider LMA-solution, VO-solution
and SMA-solution. For the LMA-solution ($\Delta m^2_{12}\simeq 3\cdot 10^{-5}$
$eV^2$) and the VO-solution ($\Delta m^2_{12}\simeq 4\cdot 10^{-10}$ $eV^2$)
we assume the following neutrino mixing matrix form:   
\begeq 
    U\simeq\left(\begin{array}{ccc} \frac{1}{\sqrt{2}}&\frac{1}{\sqrt{2}}&0\\ 
                               -\frac{1}{2} & \frac{1}{2}
                              &\frac{1}{\sqrt{2}}\\                          
                              \frac{1}{2}&-\frac{1}{2}&\frac{1}{\sqrt{2}}\\  
      \end{array}\right) 
\fineeq 
\begeq 
   M_{heavy} = \left(\begin{array}{ccc} 
    \frac{m_u^2}{2}\left(\frac{1}{m_1}+\frac{1}{m_2}\right) &   
    \frac{m_u m_c}{2\sqrt{2}}\left(\frac{1}{m_1}-\frac{1}{m_2}\right) &  
    \frac{m_u m_t}{2\sqrt{2}}\left(\frac{1}{m_2}-\frac{1}{m_1}\right) \\ 
    \frac{m_u m_c}{2\sqrt{2}}\left(\frac{1}{m_1}-\frac{1}{m_2}\right) & 
    \frac{m_c^2}{4}\left(\frac{1}{m_1}+\frac{1}{m_2}+\frac{2}{m_3}\right) &   
    \frac{m_c m_t}{4}\left(\frac{2}{m_3}-\frac{1}{m_1}-\frac{1}{m_2}\right)\\ 
    \frac{m_u m_t}{2\sqrt{2}}\left(\frac{1}{m_2}-\frac{1}{m_1}\right) & 
    \frac{m_c m_t}{4}\left(\frac{2}{m_3}-\frac{1}{m_1}-\frac{1}{m_2}\right) & 
    \frac{m_t^2}{4}\left(\frac{1}{m_1}+\frac{1}{m_2}+\frac{2}{m_3}\right)    
    \end{array}\right)\nonumber\\ 
\fineeq 
For the SMA-solution ($\Delta m^2_{12}\simeq 5\cdot 10^{-6}$
$eV^2$) we assume, instead:
\begeq   
   U\simeq\left(\begin{array}{ccc} 1& 0 &0 \\ 0 & \frac{1}{\sqrt{2}}&\frac{1}
     {\sqrt{2}}\\  
        0&-\frac{1}{\sqrt{2}} &\frac{1}{\sqrt{2}}  
       \end{array}\right)  
\fineeq  
As a consequence, resorting to (19), we have:  
\begeq  
  M_{heavy}=\left(\begin{array}{ccc} \frac{m_u^2}{m_1} &0&0\\   
     0& \frac{m_c^2}{2}\left(\frac{1}{m_2}+\frac{1}{m_3}\right)&  
     \frac{m_c m_t}{2}\left(\frac{1}{m_3}-\frac{1}{m_2}\right)\\  
    0& \frac{m_c m_t}{2}\left(\frac{1}{m_3}-\frac{1}{m_2}\right)&  
     \frac{m_t^2}{2}\left(\frac{1}{m_2}+\frac{1}{m_3}\right)  
  \end{array}\right)  
\fineeq  
The eigenvalues of these matrixes are the masses of heavy neutrinos. There is a
problem: we only know $\Delta m^2_{ij}$ from the analysis of solar
and atmospheric neutrinos ($\Delta m^2_{13}\simeq 3\cdot 10^{-3}$
$eV^2$); thus we have to consider the four possible
relationships between the light neutrino masses $m_i$:
\begin{enumerate}
  \item $m_1\ll m_2\ll m_3$ 
  \item $m_1\simeq m_2\ll m_3$ 
  \item $m_1\simeq m_2\gg m_3$
  \item $m_1\simeq m_2\simeq m_3\gg\Delta m_{13}^2$
\end{enumerate}
By numerical calculation we obtain the results reported in table
\ref{massa}.
\begin{table}[htb]  
\begin{center}   
\caption{Estimate of heavy neutrinos masses (all masses are in  GeV)}      
\label{massa}   
\bigskip      
\begin{tabular}{|c|c|c|c|c|}   
\hline    
\hline  
{\it Solutions} & Relations among $m_i$ &$M_1$&$M_2$&$M_3$ \\ \hline 
\hline      
\hline
           &$m_1\ll m_2\ll m_3$&$3\cdot 10^6$&$6\cdot 10^{10}$&$\gg 10^{15}$
           \\\cline{2-5}            
           &$m_1\simeq m_2\ll m_3$&$2\cdot 10^5-2\cdot
           10^6$&$4\cdot10^{10}$&$2\cdot 10^{14}-2\cdot 10^{15}$
           \\\cline{2-5}      
\raisebox{2.0ex}[3mm]{{\it LMA}}&$m_1\simeq m_2\gg m_3$&$2\cdot
           10^5$&$5\cdot10^{10}$&$\gg2\cdot10^{14}$ 
           \\\cline{2-5}        
           &$m_1\simeq m_2\simeq m_3$&$\gg
           4\cdot10^3$&$\gg4\cdot10^9$&$\gg10^{13}$\\ 
\hline  
\hline          
           &$m_1\ll m_2\ll m_3$&$\gg4\cdot 10^6$&$6\cdot 10^{10}$&$7\cdot
           10^{15}$ 
           \\\cline{2-5} 
           &$m_1\simeq m_2\ll m_3$&$2\cdot 10^5-5\cdot
           10^6$&$6\cdot10^{10}$&$6\cdot 10^{14}-8\cdot 10^{15}$
           \\\cline{2-5}
 \raisebox{2.0ex}[3mm]{{\it SMA}}&$m_1\simeq m_2\gg m_3$&$2\cdot
           10^5$&$6\cdot10^{10}$&$\gg10^{14}$ 
           \\\cline{2-5}         
           &$m_1\simeq
           m_2\simeq m_3$&$\gg 4\cdot10^3$&$\gg4\cdot10^9$&$\gg10^{13}$\\
\hline  
\hline 
           &$m_1\ll m_2\ll m_3$&$9\cdot 10^8$&$5\cdot 10^{10}$&$\gg3\cdot 10^{17}$
           \\\cline{2-5}
           &$m_1\simeq m_2\ll m_3$&$2\cdot 10^5-5\cdot
           10^8$&$6\cdot10^{10}$&$2\cdot 10^{14}-5\cdot 10^{17}$
           \\\cline{2-5}
\raisebox{2.0ex}[3mm]{{\it VO}}&$m_1\simeq m_2\gg m_3$&$2\cdot
           10^5$&$5\cdot10^{10}$&$\gg2\cdot10^{14}$ 
           \\\cline{2-5}        
           &$m_1\simeq m_2\simeq m_3$&$\gg
           4\cdot10^3$&$\gg4\cdot10^9$&$\gg10^{13}$\\ 
\hline        
\hline   
\end{tabular}   
\end{center}   
\end{table}   

Only one of the heavy neutrino masses is of the order of magnitude of the
scale of the $SU(5)\times U(1)$ symmetry limit. This does not contradict our
assumptions: a {\it democratic} mass matrix with all matrix elements of order
$\sim 10^{15} GeV$ would give two vanishing diagonal masses and only one mass
of order $10^{15} GeV$.

\section{Radiative lepton number violating decays}

Of all heavy neutrino decay modes, for example 
 $\nu_L\phi$ or $e^- W^+$, the Higgs channel should dominate.  We calculated
\cite{tesi} the decay rate for the process $N\rightarrow e^- W^+$ obtaining a
lifetime of order of $10^{-8}s$.    

Since the heavy neutrinos cannot be produced in accelerators and their
lifetime is so low, it will be very difficult to observe them, except,
indirectly, through  their contribution to radiative processes such as the
$\mu\rightarrow e\gamma$ decay.

\begin{figure}[ht]    \begin{center}  
 \bigskip\bigskip\bigskip\bigskip\bigskip  
  \bigskip\bigskip   
\begin{fmffile}{decgamma}       
\begin{fmfgraph*}(200,20) \fmfpen{thick}    
  \fmfleft{i1}    
  \fmfright{o1}
  \fmfforce{.6w,1.9h}{t1}
  \fmfforce{.85w,3.2h}{t2}
  \fmf{photon,label=$\gamma$}{t1,t2}
  \fmf{fermion,label=$\mu$}{i1,v1}    
  \fmf{fermion,label=$\nu_i$}{v1,v2}    
  \fmf{fermion,label=$e$}{v2,o1}
  \fmf{photon,label=$W$,left,tension=.005}{v1,v2} 
\end{fmfgraph*}      
\end{fmffile}    
\end{center}    
\bigskip   
\caption{The one-loop diagram for the $\mu\rightarrow e\gamma$ decay}   
\label{gamma}   
\end{figure}
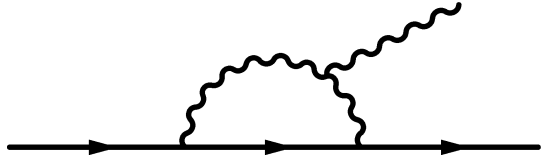

The $\mu\rightarrow e\gamma$, $\tau\rightarrow \mu\gamma$ and
$\tau\rightarrow e\gamma$ decays violate  lepton number conservation and
 are not allowed in the standard model. The Majorana mass terms,
violate lepton number and these decays could be allowed
(figure \ref{gamma}). 

The neutrino weak eigenstates are a superposition of three light
Majorana neutrinos and three heavy Majorana neutrinos, thus the branching
ratios, calculated by Cheng and Li \cite{decagamma}, are:
\begeq
  B(\mu\rightarrow e\gamma)=\frac{\Gamma(\mu\rightarrow e\gamma)}{\Gamma(
  \mu\rightarrow e\nu\overline{\nu})}=\frac{3\alpha}{8\pi}
  \left|U_{\mu i}U_{ei}\left(\frac{m_i}{M_i}\right)^2\right|^2
\fineeq
where $m_i$ are the neutrino Dirac masses and $M_i$ are the heavy neutrino
masses. For our values of heavy neutrino masses we get: $B\le 10^{-30}$ 
for $\mu\rightarrow e\gamma$, much smaller than the present limit
($B\le10^{-11})$. The branching ratios for the processes
$\tau\rightarrow \mu\gamma$ and $\tau\rightarrow e\gamma$ turn out to be
respectively $B\le 10^{-30}$ and $B\le 10^{-31}$. 

\section{Conclusion}
After introducing   the see-saw
mechanism, we have analysed this mechanism assuming the SO(10) model and
that the Dirac masses of fermions are generated by a Higgs decuplet. Even if in
this way we obtain the bad relationship $M_d=M_l$ we use it to estimate the
masses of heavy neutrinos. We have found that not all  heavy Majorana
neutrino masses are equal and that
their mass matrix can not be equal to that of up quarks. We 
tried to estimate these masses, finding that, in most cases
considered, they are: $M_1= 10^5-10^6$ $GeV$, $M_2= 10^9-10^{10}$ $GeV$ and
$M_3=  10^{13}-10^{16}$ $GeV$. We also calculated that their lifetime is too
short to be present in cosmic rays and finally we analysed in which way the
heavy neutrino can  affect the $\mu\rightarrow e\gamma$ decay (and 
$\tau\rightarrow \mu\gamma$,  $\tau\rightarrow e\gamma$)  
estimating a rate which is far below the present limit.

\end{document}